\documentstyle[11pt,amsfonts]{article}

\newcommand{\be}{\begin{equation}}
\newcommand{\ee}{\end{equation}}
\newcommand{\bea}{\begin{array}}
\newcommand{\ea}{\end{array}}
\newcommand{\beqa}{\begin{eqnarray}}
\newcommand{\eeqa}{\end{eqnarray}}
\newcommand{\bean}{\begin{eqnarray*}}
\newcommand{\eean}{\end{eqnarray*}}

\def\up#1{\leavevmode \raise.16ex\hbox{#1}}

\newcommand{\gapproxeq}{\lower
 .7ex\hbox{$\;\stackrel{\textstyle >}{\sim}\;$}}
\newcommand{\lapproxeq}{\lower .7ex\hbox{$\;\stackrel
{\textstyle <}{\sim}\;$}}

\newcounter{appendice}

\def\thebibliography#1{{\bf REFERENCES\markboth
 {REFERENCES}{REFERENCES}}\list
 {[\arabic{enumi}]}{\settowidth\labelwidth{[#1]}\leftmargin\labelwidth
 \advance\leftmargin\labelsep
 \usecounter{enumi}}
 \def\newblock{\hskip .11em plus .33em minus -.07em}
 \sloppy
 \sfcode`\.=1000\relax}

\def\BI{{\rm 1\!l}}

\begin{document}

\centerline{ \LARGE  Gauge Theory of the Star Product }

\vskip 2cm

\centerline{ {\sc    A. Pinzul$^{a}$ and A. Stern$^{b}$ }  }

\vskip 1cm
\begin{center}
{\it a)Instituto de F\'{i}sica, Universidade de S\~{a}o Paulo\\
C.P. 66318, S˜ao Paulo, SP, 05315-970, Brazil\\}
{\it b) Department of Physics, University of Alabama,\\
Tuscaloosa, Alabama 35487, USA}

\end{center}

\vskip 2cm

\vspace*{5mm}

\normalsize
\centerline{\bf ABSTRACT}
\vspace*{5mm}

The choice of a star product realization  for  noncommutative field
theory can be regarded as a gauge choice in the space of all equivalent
star products.   With the goal of having a gauge invariant treatment,
we develop tools, such as  integration measures and  covariant derivatives on
this space.  The covariant derivative can be expressed in terms of
connections in the usual way giving rise to new degrees of freedom for
noncommutative theories.

\newpage
\scrollmode
\section{Introduction}
\setcounter{equation}{0}

Deformation quantization\cite{Sternheimer}  replaces the commutative algebra of functions
on a Poisson manifold with a non-commutative algebra, where
multiplication for the latter is given by some associative star product.   Kontsevich 
\cite{Kontsevich:1997vb} showed  that this program can be carried out for
any smooth Poisson manifold.  An explicit construction for the star
product (the Kontsevich star product) was given in
\cite{Kontsevich:1997vb} which was  completely determined by
 the Poisson bi-vector $\alpha$.  The
Kontsevich star product  
belongs to a very large equivalence class $\{\star\}_{\alpha}$ of star
products.  The different star     products
in  $\{\star\}_{\alpha}$ are related by gauge transformations,  where
the gauge group  ${\cal
  G}_\alpha$ is generated by all differential operators.  ${\cal
  G}_\alpha$
 also  includes  transformations, such as standard
noncommutative $U(1)$  gauge transformations, which leave the star product invariant.\cite{Jurco:2000fb},\cite{Jurco:2001my}

 Other ingredients, in addition to the non-commutative algebra, are needed in order to write down
field theories on the non-commutative spaces.  Among them are the trace
and derivative.  Concerning the former, theorems have been given which
show   the existence of the trace for the
deformation of a symplectic manifold \cite{Gutt} and, more generally,
for any regular  Poisson manifold for which a Poisson trace exists.\cite{Felder:2000nc},\cite{Dolgushev}   It was shown,
for example,  that  the usual integral with the commutative measure
satisfies the necessary  conditions for a trace when the topology of the manifold is
${\mathbb{R}}^{2d}$ and one restricts to the Kontsevich star.
In another example, corrections were computed   in \cite{Stern:2006zt}  to the commutative
measure for a star product constructed from deformed
 coherent states\cite{Alexanian:2000uz}.  Concerning the
derivative, much is known  for
the case of constant non-commutativity, where the Kontsevich star 
reduces to the familiar Groenewold-Moyal star\cite{Groenewold:1946kp},\cite{Moyal:1949sk}.   For that case,
the standard partial derivative can be realized as an inner derivative on the algebra.
This is not true  for
 non-constant non-commutativity where the star product is position dependent.
Derivatives have nevertheless  been defined in the general case, after
specializing to the Kontsevich star
product.\cite{Behr:2003qc}

As most previous treatments of noncommutative field theory have relied heavily  on one
particular star product in  $\{\star\}_{\alpha}$, namely  the
Kontsevich star product, it is of interest to search for  gauge
invariant approaches, where here the gauge group is ${\cal G}_\alpha$.  This is addressed in the current article.   One
 approach is to simply  map the  noncommutative field theory written with the
 Kontsevich star to a  noncommutative field theory associated with an arbitrary
 star product in the equivalence class, resulting in
 no  new physical degrees of
 freedom.  Alternatively, one can
 introduce the notion of  covariance with respect to  ${\cal
   G}_\alpha$, where  functions  $\{f,g,...\}$ and their star products
 transform in the same manner, and are hence covariant.  In addition,
 one can define the trace of the functions to be gauge invariant.  Like in
Yang-Mills theories, one can then also
introduce  a covariant derivative, now associated with gauge transformations
between different star products in  $\{\star\}_{\alpha}$, where the
covariant derivative of  functions  $\{f,g,...\}$  transforms in the same manner as
the  functions  $\{f,g,...\}$.   In such an approach, which is what we follow here, 
one thereby obtains  new degrees of freedom associated with the
connections.\footnote{ Only after
restricting the connection to a certain pure gauge, will the covariant
derivative  be gauge equivalent to the derivative
\cite{Behr:2003qc}  written for the Kontsevich star.}   As ${\cal G}_\alpha$
 is an infinite-dimensional extension  of the noncommutative $U(1)$ gauge group, there are in principle
an infinite number of such degrees of freedom, which contains the standard
noncommutative $U(1)$ gauge degrees of freedom.  It then becomes
possible to consider an infinite-dimensional extension of
noncommutative $U(1)$ gauge theory, with the
dynamics of gauge fields and matter fields written on the entire equivalence class $\{\star\}_{\alpha}$.

The plan of the paper is as follows. In section 2 we introduce the
integration measure,
covariant derivative, connection and curvature for the special case
where $\{\star\}_{\alpha}$ contains the Groenewold-Moyal star
product, while the generalization to an arbitrary equivalence class is
given in section 3. Arbitrary star products in
$\{\star\}_{\alpha}$ can be 
expanded in the noncommutative parameter, which we denote by $\hbar$,
and each order can be expressed in terms of an infinite number of
bi-differential operators.   Furthermore, ${\cal G}_\alpha$ is generated by an
infinite number of differential operators at each order in $\hbar$.
For practical purposes, we examine a restricted gauge group in section
4 which is generated by a finite number of differential operators at
each order in $\hbar$.  We can then write down explicit formulae for
components of the connection, curvature and field equations.   We
summarize the results and indicate possible future developments in section 5.

\section{Gauging the Groenewold-Moyal star product} 
\setcounter{equation}{0}

We first review well known facts about the  Groenewold-Moyal star product.
Here the Poisson bi-vector  on
 ${\mathbb{R}}^{2d}$   coordinatized by $x^\mu,\;\mu=1,2,...,2d$, is
\be  \theta^{\mu\nu}\overleftarrow{
  \partial_\mu}\;\overrightarrow{ \partial_\nu} \;,\ee
where $\theta^{\mu\nu}=-\theta^{\nu\mu}$ are constants  on
 ${\mathbb{R}}^{2d}$.
  $\overleftarrow{
  \partial_\mu}$
 and $\overrightarrow{ \partial_\mu} $  are left and right
derivatives ${ \partial_\mu}= { \frac\partial { \partial x^\mu }}$,
 respectively.  Constant non-commutativity results after deformation
 quantization.  
Denote by  ${\cal A}_{\theta}$ the noncommutative algebra of functions $f_0,g_0,...$ on
 ${\mathbb{R}}^{2d}$ with multiplication given by the Groenewold-Moyal
 star product $\star_\theta $ \cite{Groenewold:1946kp},\cite{Moyal:1949sk}.    
  $\star_\theta $ is the bi-differential operator 
\be \star_\theta = \exp\;\biggl\{ \frac {i\hbar}2 \theta^{\mu\nu}\overleftarrow{
  \partial_\mu}\;\overrightarrow{ \partial_\nu} \biggr\} \;. \ee
Then at lowest order in $\hbar $, the star commutator of functions reduces to their
Poisson bracket
\be  [f_0, g_0]_{\star_\theta}\equiv  f_0\star_\theta g_0-
g_0\star_\theta f_0=i\hbar \{f_0,g_0 \} \;+\;{\cal O}(\hbar^3)\;,\label{hblmtfsc} \ee
where $ \{f_0,g_0 \}=  f_0 \; \theta^{\mu\nu}\overleftarrow{
  \partial_\mu}\;\overrightarrow{ \partial_\nu}\; g_0$.
 The
 derivative $\partial_\mu$ satisfies the usual Leibniz rule when acting
 on  the Groenewold-Moyal
 star product of two functions. 
 Using the
 standard measure $ d^{2d}x$ on  ${\mathbb{R}}^{2d}$, the integral
 serves as a trace for  ${\cal A}_{\theta}$.   Moreover, the integral
 of  the Groenewold-Moyal
 star product of two functions  can be replaced with the
 integral of the pointwise product of the two functions, provided these functions
 vanish sufficiently rapidly at infinity 
\be \int d^{2d}x\; f_0\star_\theta g_0 =
 \int d^{2d}x\; f_0\;g_0\;,\label{streqlpntws} \ee
from which the trace property easily follows, 
\be \int d^{2d}x\; [f_0, g_0]_{\star_\theta} = 0\;. \ee 

The Groenewold-Moyal star product  $\star_\theta $ is an element of the equivalence
 class of star products $\{\star\}_{\theta}$.   The equivalence class is
 generated from the set of all invertible operators $T$
 of the form \be T= \BI + \sum_{k=1}^\infty\hbar^k\;T_k
\label{tauxpns}\;,\ee  where $T_k$ are arbitrary differential operators.
 Under the action of $T$, functions   $f_0,g_0,...$ are mapped to
\be  f=T(f_0)\;,\quad g=T(g_0)\;,... \;,\label{ff0gg0} \ee while 
 $\star_\theta$ is mapped to another associative star product $\star \in\{\star\}_{\theta}$, such that\cite{Kontsevich:1997vb}
\be    f\star g= T( f_0\star_{\theta}  g_0) \label{eqvcndtn}\;.  \ee 
The new star commutator has the same $\hbar\rightarrow 0 $ limit as in (\ref{hblmtfsc}),
\be  [f, g]_{\star}\equiv  f\star g-
g\star f=i\hbar \{f,g \} \;+\;{\cal O}(\hbar^2)\;. \ee
As before, the
 integral  serves as the
 trace.  However, the measure associated with the star product $\star$ is, in
 general, no longer $ d^{2d}x$.  Call the transformed measure
 $d\mu_\star\;$.   Invariance of the trace implies
\be \int d\mu_\star\; f = \int d^{2d}x\;  f_0  \;,\label{trofmsrfrmk}
 \ee for functions $f_0$ vanishing sufficiently rapidly at infinity.
From (\ref{tauxpns}), $d\mu_\star$ can differ from the flat measure  $
d^{2d}x$ at order $\hbar$.   (An explicit expression for  $d\mu_\star$ for
 restricted gauge transformations is given in
 sec. 4.3.)
 The analogue of (\ref{streqlpntws}) does not hold for  an arbitrary  $\star$ in the equivalence class ; i.e. $ \int d\mu_\star \;
 f\star g\ne  \int d\mu_\star \;
 f\; g $.  However, the trace property
 easily follows from (\ref{eqvcndtn}) and (\ref{trofmsrfrmk})
\beqa \int d\mu_\star \;[ f,g]_\star& =&\int d\mu_\star \;T([
 f_0,g_0]_{\star_\theta})\cr &=& \int d^{2d}x\;[
 f_0,g_0]_{\star_\theta} = 0 \;. \eeqa

Subsequent  transformations can be performed  to map between any two star
products in $\{\star\}_{\theta}$.  Say that  $\star'$ is obtained from
$\star$ using  invertible  operator $\Lambda$, which we
assume to have a form analogous to (\ref{tauxpns}),
\be \Lambda= \BI + \sum_{k=1}^\infty\hbar^k\;\Lambda_k
\label{lmbdauxpns}\;,\ee  where $\Lambda_k$ are arbitrary differential
operators.
  Then
functions, as well as star products of functions, transform `covariantly':
\beqa  f \;\;
 \rightarrow\quad f'\quad&=& \Lambda( f)\;,\cr & &\cr    f\star  g
 \rightarrow  f' \star' g' & =& \Lambda( f\star  g)\;.
\label{sosubggtrnsone}
\eeqa
In  order that the trace be an invariant for $\{\star\}_{\theta}$, the measure should, in general,  transform, $d\mu_\star\rightarrow
d\mu_{\star'}$, such that 
\be \int d\mu_\star\; f =\int d\mu_{\star'}\; f'\;,\label{nvrsftrc} \ee 
for all functions $f$ that 
 vanish sufficiently rapidly at infinity.  $d\mu_\star$ and $
d\mu_{\star'}$ correspond  to the flat measure  $
d^{2d}x$ at zeroth order in $\hbar$, while from (\ref{lmbdauxpns}),
they, in general,  differ at order $\hbar$.   Since the measure transforms
 nontrivially  for  general $\Lambda$, gauge transformations cannot be considered to be  
 internal transformations beyond zeroth order in $\hbar$.

  There exists a subset of transformations
 (\ref{lmbdauxpns})  which leaves the star product and  the measure invariant, i.e.,
 $\star'=\star$ and $d\mu_{\star'}=d\mu_\star\;$.   This is the case for   inner automorphisms \cite{Jurco:2000fb},\cite{Jurco:2001my} $\Lambda =
\hat \Lambda_\lambda$, parametrized by  functions $\lambda$ on  ${\mathbb{R}}^{2d}$,  where   \be
\hat \Lambda_\lambda(f)=\lambda \star f\star\lambda^{-1}_\star\;,\label{inrot}\ee 
 and $\lambda \star \lambda^{-1}_\star=1.$  
  It is not surprising that  general gauge transformations given by (\ref{sosubggtrnsone})
 are not internal beyond zeroth order in $\hbar$, because the same is true for the subset of inner
 automorphisms, even though the measure is invariant under the latter.  For the case of the  Groenewold-Moyal
 star, the  inner
 automorphisms are known to contain (global) translations.\cite{Szabo:2006wx}  Here we need that
 $[\theta_{\mu\nu}]$ has an inverse:
\be e_\star^{\;-i\theta^{-1}_{\rho\sigma}c^\rho x^\sigma}\; \star_\theta
\; x^\mu\; \star_\theta \; e_\star^{\;i\theta^{-1}_{\rho\sigma}c^\rho x^\sigma}
 = x^\mu +\hbar c^\mu\;, \ee where $e_\star^{\;f} = 1 + f +\frac12
 f\star f + \frac 1{3!} f\star f\star f + \cdot\cdot\cdot\;.$  So for
 $c^\mu$ of zeroth order in $\hbar$, one gets translations of order $\hbar$.

We denote the   derivative associated with any
$\star\in\{\star\}_{\theta}$ by
$D[A]_\mu$, and require that it is covariant under the  gauge
transformations (\ref{sosubggtrnsone}),
 \beqa
D[A]_\mu f
& \rightarrow &D [A']_\mu f'   = \Lambda(D[A]_\mu f)
\;,\label{cvdrvtvdef} \eeqa or
\be D [A']_\mu \;\Lambda   = \Lambda\; D[A]_\mu\;. \ee
Since $\Lambda$ is a  differential operator of arbitrary order, so in general should be $
D[A]_\mu$.   As usual, let us write the covariant derivative in terms of
potentials  $A_\mu$, which we  expand according to
\be A_\mu=  \sum_{k=1}^\infty\hbar^k\;A_{k,\mu}
\label{almbdauxpns}\;,\ee  where $A_{k,\mu}$ are differential
operators.\footnote{Derivative-valued gauge fields
  have been considered previously in  different
  contexts.\cite{Jurco:2000fb},\cite{Jurco:2001my},\cite{Dimitrijevic:2003pn}}  If we require $
D[A]_\mu$ to reduce to the standard derivative in the absence of the
potentials, then we can  write the usual expression 
\be  D[A]_\mu=\partial_\mu + A_\mu\;,\ee and the potentials $ A_\mu$
 gauge transforms as
\be A_\mu\rightarrow A'_\mu = \Lambda [\partial_\mu,\Lambda^{-1} ] +
\Lambda A_\mu\Lambda^{-1}\label{gtfdvptnls}\;. \ee

Derivative-valued field strengths \be F_{\mu\nu}=  \sum_{k=1}^\infty\hbar^k\;F_{k,\mu\nu}
\label{drvldfs}\;,\ee  where $F_{k,\mu\nu}$ are arbitrary differential
operators,  can  also be introduced
\be F_{\mu\nu}= [ D[A]_\mu,D[A]_\nu] =[\partial_\mu,
A_\nu]-[\partial_\nu, A_\mu]+ [A_\mu,A_\nu]\;.\label{fsntcs} \ee They
satisfy the Bianchi identity
\be [D[A]_\rho,F_{\mu\nu}]+ [D[A]_\mu,F_{\nu\rho}]+
[D[A]_\nu,F_{\rho\mu}]=0\;,\ee and gauge
transform according to  \beqa
F_{\mu\nu} 
& \rightarrow &F_{\mu\nu}'   = \Lambda F_{\mu\nu} \Lambda^{-1}\label{ggtrnofF}
\;. \eeqa  For the special case where   $A_\mu$ is the pure gauge $A_\mu =T
[\partial_\mu,T^{-1} ]$, $ D[A]_\mu$ satisfies the usual Leibniz rule when acting
 on  the
 star product $\star$ of two functions.  This, however, is not true
 for arbitrary connections $A_\mu$.

$U(1)$ gauge theory on the noncommutative plane is contained in this
system.   Here  we write $A_\mu=\hat A_\mu$  acting on
  functions $f$ [which gauge  transform as inner automorphisms
(\ref{inrot})] according to  $\hat A_\mu(f)=[ a_\mu, f]_{ \star_\theta}$, where $a_\mu$ are the noncommutative
  $U(1)$ potentials.  So the derivative-valued potentials acting on
  covariant functions can be
  written as\footnote{Alternatively, on can define the action of $\hat
   A$ on fields  $\phi_{fund}$ in the  fundamental representation.
Such fields are not covariant, in that they do not gauge
transform according to (\ref{sosubggtrnsone}) with
$\Lambda=\hat\Lambda_\lambda$, but rather with the left action 
$\phi_{fund} \rightarrow\phi'_{fund}=\lambda \star \phi_{fund}$.
On such fields one has  $$\hat A_\mu =
a_\mu \exp \Bigl\{   \frac {i\hbar}2
\theta^{\rho\sigma}\overleftarrow{\partial_\rho}\;\overrightarrow{\partial_\sigma}\Bigr\}\;.$$
}
\be \hat A_\mu = 2i\; a_\mu \sin \Bigl\{   \frac {\hbar}2 \theta^{\rho\sigma}\overleftarrow{
  \partial_\rho}\;\overrightarrow{ \partial_\sigma}\Bigr\}\;. \ee Upon
  restricting  to the  Groenewold-Moyal star   $\star_\theta $,
  (\ref{inrot}) leads to recover the
  usual noncommutative $U(1)$ gauge transformations for $a_\mu$,
\be a_\mu\rightarrow a'_\mu = \lambda\star_\theta a_\mu\star_\theta\lambda_\star^{-1}
  - \partial_\mu\lambda\star_\theta\lambda_\star^{-1}\;. \ee
From (\ref{fsntcs}), the field strength operators $F_{\mu\nu}=\hat F_{\mu\nu}$
  acting on a function $\phi$ is  $\hat F_{\mu\nu}(\phi)=[ f_{\mu\nu},  \phi]_{ \star_\theta }$, where $ f_{\mu\nu}$ is the noncommutative
  $U(1)$ field strength tensor $ f_{\mu\nu}=\partial_\mu a_\nu-
  \partial_\nu a_\mu +[ a_\mu, a_\nu]_{\star_\theta} $.   Thus  \be \hat  F_{\mu\nu} = 2i\; f_{\mu\nu} \sin \Bigl\{   \frac {\hbar}2 \theta^{\rho\sigma}\overleftarrow{
  \partial_\rho}\;\overrightarrow{ \partial_\sigma}\Bigr\}\;,\label{ncuonef} \ee and upon
  restricting $\Lambda$ in (\ref{ggtrnofF}) to (\ref{inrot}),
  $f_{\mu\nu}$ gauge transform as inner automorphisms,
$ f_{\mu\nu}\rightarrow f'_{\mu\nu} = \lambda\star_\theta f_{\mu\nu}\star_\theta\lambda_\star^{-1}$.

Field  theory actions can now be written down which are
invariants for the equivalence class  $\{\star\}_{\theta}$.  If we
assume the field
$\phi$  on  ${\mathbb{R}}^{2d}$ transforms  
covariantly with respect to the above gauge transformations, $ \phi
 \rightarrow \phi'= \Lambda( \phi)$, then an invariant action is
\be S_{\phi,A} = \frac 12 \int d\mu_\star\;\;\eta^{\mu\nu}\;D[A]_\mu\phi\;\star
\;D[A]_\nu\phi\;,\label{gnvactnfrmtr} \ee where $\eta^{\mu\nu}$ is the flat metric.  It is equivalent to the commutative action for a free
massless scalar field after restricting $A_\mu$ to  the pure gauge $A_\mu =T
[\partial_\mu,T^{-1} ]$ and making the field redefinition from
$\phi=T(\phi_0)$ to $\phi_0$.  This is since then $D[A]_\mu\phi =
T\partial_\mu\phi_0$, and  we can  re-express the action in terms of the
Groenewold-Moyal star product and use (\ref{streqlpntws}). 

 More
exciting is the possibility of writing down a kinetic term for
$A_\mu$.   This would require a trace over  the operator-valued
fields.  A possible candidate is the Wodzicki residue\cite{Wod}.
Alternatively, one can adopt the usual Yang-Mills form for the field
equations:
\be [D[A]_\mu,F^{\mu\nu}] = J^\nu \;.\label{ymteq}\ee  The right hand side
represents a matter current source which can be expanded \be J_{\mu}=  \sum_{k=1}^\infty\hbar^k\;J_{k,\mu}
\label{drvldcrnt}\;,\ee and which  gauge transforms as the field
strengths $F_{\mu\nu}$, \be J^\mu\rightarrow
J'^\mu= \Lambda J^\mu\Lambda^{-1}\;.\label{gtrnofJ} \ee  Moreover, it must be covariantly
conserved,
\be  [D[A]_\mu, J^\mu]=0 \;.\label{nftcnsrvlws}\ee
Since $J^\mu$ takes values in an infinite dimensional vector space,
(\ref{nftcnsrvlws}) then corresponds to infinitely many conservation laws.

\section{Generalization to the Kontsevich star product}
\setcounter{equation}{0}

Now we go to the case of a general Poisson bi-vector
\be \overleftarrow{
  \partial_\mu}\; \alpha^{\mu\nu}\;\overrightarrow{ \partial_\nu}\;,\label{pbvvtr} \ee
  where  $\alpha^{\mu\nu}=
 -\alpha^{\nu\mu},\; \mu,\nu = 1,2,...,2d$  are functions on an   open
 subset $M^{2d}$ of ${\mathbb{R}}^{2d}$.
Corresponding star products can be given in terms of 
 series expansions in the non-commutativity parameter $\hbar$, where
 the terms in the expansions are
 bi-differential operators  $B_n,\;n=1,2,3...$,  \be f\star g = fg
 +\sum_{n=1}^\infty \hbar^n B_n(f,g)\;.\label{gnrlfrmnqc} \ee
In the Kontsevich construction of the star product \cite{Kontsevich:1997vb}, which we
denote using $\star_\alpha$, $B_1$ is proportional to the
 Poisson bi-vector field.  Acting between
 functions $f_0$ and $g_0$, $\star_\alpha$ is, up to  second order
 in $\hbar$, given by\footnote{Third order terms were computed  in \cite{Zotov:2000ec}.}
\beqa f_0\star_\alpha g_0 &=& f_0g_0\;
    +\;\frac{i\hbar}2 \alpha^{\mu\nu}\partial_\mu f_0\partial_\nu g_0 \;  -\;\frac
    {\hbar^2}8 \alpha^{\mu\nu} \alpha^{\rho\sigma}
    \partial_{\mu,\rho} f_0\partial_{\nu,\sigma}
    g_0
\cr & &\cr& & - \;\frac
    {\hbar^2}{12}
    \alpha^{\mu\nu}\partial_\nu\alpha^{\rho\sigma}(\partial_{\mu,\rho} f_0\partial_\sigma
    g_0 - \partial_\rho f_0\partial_{\mu,\sigma}
    g_0)
  \;+\;{\cal O}(\hbar^3) \;,  \label{strknstvh} \eeqa
where $\partial_{\mu,\nu,...,\rho}=\frac\partial{\partial
 x^\mu}\frac\partial{\partial
 x^\nu}\cdot\cdot\cdot\frac\partial{\partial x^\rho}$.
 The  Poisson bracket is again recovered at   lowest order in $\hbar$ from the star commutator 
 \beqa  [f_0,g_0]_{\star_\alpha}\equiv  f_0\star_\alpha  g_0 - 
 g_0\star_\alpha f_0
& =&i\hbar \{f_0,g_0 \} \;+\;{\cal O}(\hbar^3)  \label{cmtvlmt}\;,
 \eeqa where $ \{f_0,g_0 \} =  f_0\; \overleftarrow{
  \partial_\mu}\; \alpha^{\mu\nu}\;\overrightarrow{ \partial_\nu} \;g_0$.

The integral with measure
$ d\mu_{0} =  d^{2d}x\;\Omega_0(x)$
 can serve as a   trace for a  star product associated with the
 Poisson bi-vector (\ref{pbvvtr}), provided that $\Omega_0(x)$ satisfies\cite{Gutt}\cite{Felder:2000nc},\cite{Dolgushev}
\be \partial_\mu(\Omega_0 \alpha^{\mu\nu}) = 0 \label{msrfctn}\ee  From
this relation  the cyclicity property  easily follows  at first order in $\hbar$,
\be \int d\mu_{0} \; [f_0,g_0]_{\star_{\alpha}}
={\cal O}(\hbar^2)\;,\label{trcprtK} \ee
 provided 
   functions $f_0$ and $g_0$ vanish sufficiently rapidly at
 infinity. For the special case of symplectic manifolds,
 $\Omega_0$ is proportional to $|\det\alpha|^{-1/2}$.  More generally, it is known\cite{Felder:2000nc} that there   exists a star
 product $\star_0$, that is gauge equivalent to $\star_\alpha$, for which
the cyclicity property is guaranteed to all orders in $\hbar$ using measure $ d\mu_{0}$.  Call
 $T_\alpha$ the map from  $\star_\alpha$ to $\star_0$, and  define a measure $ d\mu_\alpha=d^{2d}x\;\Omega(x)$  associated with star
 product  $\star_\alpha$ such that  
\be \int d\mu_\alpha f_0 = \int d\mu_0 \;T_\alpha(f_0) \;,\ee
for all $f_0$ that vanish  sufficiently rapidly at
 infinity.  Thus
\be \int d\mu_{\alpha} \; [f_0,g_0]_{\star_{\alpha}}
=0\;,\label{trcprtK} \ee and so $\int d\mu_{\alpha}$ serves as a trace
for the star product $\star_\alpha$.

Derivations  $\delta_X^\alpha$  can be defined for the star product $\star_{\alpha}$
satisfying the standard Leibniz rule \be \delta_X^\alpha ( f_0\star_{\alpha} g_0) = \delta_X^\alpha  f_0\star_{\alpha} g_0 +
f_0\star_{\alpha} \delta_X^\alpha  g_0\;, \label{uslLrl} \ee  provided the Lie derivative  $ {\cal L}_X $ of  $\alpha$  vanishes,
\be  {\cal L}_X \alpha^{\mu\nu}=  X^\rho \partial_\rho \alpha^{\mu\nu}  -
\alpha^{\mu\rho}\partial_\rho X^\nu +\alpha^{\nu\rho}\partial_\rho
X^\mu= 0 \;,\;\label{sncndtn}\ee for some vectors  $X=
X^\mu\partial_\mu$.  This is the same condition that is
needed for $X$ to be a derivation of the Poisson bracket; i.e., $X\{f_0,g_0\}= \{Xf_0,g_0\}+ \{f_0,Xg_0\}$,
and it also corresponds to a vanishing Schouten-Nijenhuis bracket of
$X$ with the Poisson bivector.   An expansion for  $\delta_X^\alpha $  was given in
\cite{Behr:2003qc}.
Up to second order in $\hbar$, $ \delta_X^{\alpha}$ was found to be
\be  \delta_X^{\alpha} = X^\mu\partial_\mu\; +\;\frac{\hbar^2}{12}
\alpha^{\mu\nu}\partial_\nu\alpha^{\rho\sigma}\partial_{\mu,\sigma}
X^\lambda\; \partial_{\rho,\lambda} \; -\;\frac{\hbar^2}{24}
\alpha^{\mu\nu}\alpha^{\rho\sigma}\partial_{\mu,\rho}
X^\lambda\; \partial_{\nu,\sigma,\lambda}  \;  + \;{\cal O}(\hbar^3)
\;.\label{dltXalpa}\ee   The commutator of any two such derivatives $
\delta_X^{\alpha}$ and $ \delta_Y^{\alpha}$ is
 nonvanishing, with the zeroth order being the Lie bracket, $ [ \delta_X^{\alpha}, \delta_Y^{\alpha}] = {\cal L}_X Y  \;  + \;{\cal O}(\hbar^2)$.

As with the Groenewold-Moyal star product,     $\star_\alpha $ belongs
 to  an equivalence
 class of star products which we denote as $\{\star\}_{\alpha}$. 
 The equivalence class is once again
 generated from the set of all invertible differential operators $T$,
 mapping functions  $f_0,g_0,...$ to $f,g,...$ in (\ref{ff0gg0}), and
the  star product 
 $\star_\alpha$ to $\star$, whose general form is given by (\ref{gnrlfrmnqc}), with
\be    f\star g= T( f_0\star_{\alpha}  g_0) \label{eqvcndtnalf}\;.  \ee
 The measure $d\mu_\star$ associated with the the star product $\star$
 is related to $ d\mu_{\alpha}$ by
\be \int d\mu_\star\; f = \int  d\mu_{\alpha}\;  f_0  \;,\label{trofmsrfrmk}
 \ee for functions $f_0$ vanishing sufficiently rapidly at infinity.
Subsequent  transformations can again be performed  to map between any two star
products in the equivalence class
 $\{\star\}_{\alpha}$ given by (\ref{sosubggtrnsone}).  The
 corresponding measures are related by (\ref{nvrsftrc}).  For
 covariant derivatives we  again need (\ref{cvdrvtvdef}).  Now say
 that the covariant derivative $
D[A]_X$  reduces to  $ \delta_X^{\alpha}$ in the absence of the
potentials, as is  the case for 
\be  D[A]_X= \delta_X^{\alpha} + A_X\;.\ee The  derivative-valued potentials $ A_X$
 gauge transform as
\be A_X\rightarrow A'_X = \Lambda [\delta_X^{\alpha},\Lambda^{-1} ] +
\Lambda A_X\Lambda^{-1}\;. \ee  Given independent derivatives $
\delta_X^{\alpha}$ and $ \delta_Y^{\alpha}$, one can define
 field strengths
\be F_{XY}= [ D[A]_X,D[A]_Y] =[\delta_X^{\alpha}, \delta_Y^{\alpha}]+[\delta_X^{\alpha},
A_Y]-[\delta_Y^{\alpha},
A_X]+ [A_X,A_Y]\;, \ee which
 gauge
transform as in (\ref{ggtrnofF}).   For the special case where   $A_X$ is the pure gauge $A_X =T
[\delta_X^{\alpha},T^{-1} ]$, $ D[A]_X$ satisfies the usual Leibniz rule when acting
 on  the
 star product $\star$ of two functions. Gauge invariant actions
 analogous to (\ref{gnvactnfrmtr}) can be written down after
 introducing a metric over the space of vector fields $\{X,Y,...\}$.

\section{$\hbar$ expansion}
\setcounter{equation}{0}

The most general  $T_k$ and $\Lambda_k$ in (\ref{tauxpns}) and
(\ref{lmbdauxpns}), respectively, contain an infinite number
of derivatives, and map  to star products  using (\ref{eqvcndtn}), which
 then also contain
 an infinite number
of derivatives at each order in $\hbar$ beyond the zeroth
order.   Here for simplicity  we shall  restrict to operators  $T_k$ and $\Lambda_k$
 which have a finite number of derivatives.  More specifically, terms
 of order $n$ in $\hbar$ in the equivalence map will be, at most, of
 order $2n$ in derivatives.  As a result of this the
 star product $\star $, connections $A_\mu$ and curvature $F_{\mu\nu}$  can be written in terms of a finite number
of derivatives at each order in $\hbar$.

\subsection{Gauge group}

We parametrize the set of all differential operators $\{T_k=T_k^{(s)}\}$ with an infinite
number of symmetric tensors
$s=(s^{\mu_1},s^{\mu_1\mu_2},s^{\mu_1\mu_2\mu_3},...)$ which are
functions  on  ${\mathbb{R}}^{2d}$ and are polynomials in $\hbar$
starting with  order zero.
 The resulting expression for  $T^{(s)}= \BI + \sum_{k=1}^\infty\hbar^k\;T_k^{(s)}$    should be consistent
with closure
\be T^{(s')}T^{(s)}=T^{(s'')}\label{clsr}\;. \ee
A possible solution is 
\be T_k^{(s)}= \tau_{2k-1}^{(s)}+ \tau_{2k}^{(s)}\;,\qquad \tau_n^{(s)}=\frac 1{n}
s^{\mu_1\mu_2...\mu_{n}}\partial_{\mu_1,\mu_2,...,\mu_{n}}\;. \label{tausubn}
\ee  The identity corresponds to $s=0$, $T^{(0)}=\BI$.  From (\ref{clsr})
one gets
\beqa s''^\mu &=&  s^\mu +  s'^\mu + \hbar\; T^{(s')}_1 s^\mu +  {\cal
  O}(\hbar^2)\left.\matrix{\cr \cr}\right.\cr
s''^{\mu\nu} &=&  s^{\mu\nu} +  s'^{\mu\nu} + \hbar \Bigl( T^{(s')}_1
s^{\mu\nu} +   s'^{(\mu}
s^{\nu)} + s'^{\lambda(\mu}
\partial_\lambda s^{\nu )}\Bigr) +  {\cal O}(\hbar^2)\cr
s''^{\mu\nu\lambda} &=&  s^{\mu\nu\lambda} +  s'^{\mu\nu\lambda}
+\frac 14\Bigl( s'^{(\mu} s^{\nu\lambda )}+  s^{(\mu} s'^{\nu\lambda
  )}+ s'^{\eta(\mu}\partial_\eta s^{\nu\lambda)}\Bigr) +  {\cal
  O}(\hbar)\cr
s''^{\mu\nu\lambda\eta} &=&  s^{\mu\nu\lambda\eta} +  s'^{\mu\nu\lambda\eta}
+\frac 1{24} s'^{(\mu\nu} s^{\lambda\eta )} +  {\cal
  O}(\hbar)
\cr.\quad .\quad. &
&.\quad.\quad.\quad.\quad.\quad.\quad.\qquad\;,\label{sdpespps} \eeqa where $s^{(\mu\nu...\rho)}=
s^{\mu\nu...\rho}\; + $ all symmetric combinations.  Denoting the
inverse of  $T^{(s)}$ by  $  T^{(s_{inv})}={T^{(s)}}^{-1}$, we get that
\beqa s_{inv}^\mu &=&-  s^\mu + \hbar\; T^{(s)}_1 s^\mu +  {\cal O}(\hbar^2)\left.\matrix{\cr \cr}\right.\cr
s_{inv}^{\mu\nu} &=& - s^{\mu\nu} + \hbar \Bigl( T^{(s)}_1 s^{\mu\nu}  +s^{(\mu}
s^{\nu)} + s^{\lambda(\mu}
\partial_\lambda s^{\nu )}\Bigr) +  {\cal O}(\hbar^2)\cr
 s_{inv}^{\mu\nu\lambda} &=& - s^{\mu\nu\lambda}
+\frac 12 s^{(\mu} s^{\nu\lambda )}+ \frac 14 s^{\eta(\mu}\partial_\eta s^{\nu\lambda)} +  {\cal
  O}(\hbar)\cr
s_{inv}^{\mu\nu\lambda\eta} &=&  -s^{\mu\nu\lambda\eta} 
+\frac 1{24} s^{(\mu\nu} s^{\lambda\eta )} +  {\cal
  O}(\hbar)
\cr.\quad .\quad. &
&.\quad.\quad.\quad.\quad.\quad.\quad.\qquad\;\label{invsdpespps} \eeqa

\subsection{Star product}

Using the operator $T^{(s)}$  in
the equivalence relation (\ref{eqvcndtnalf}), the Kontsevich star product $\star_{\alpha}$ is
mapped to the star product  given by (\ref{gnrlfrmnqc}), with the
first two bi-differential operators $B_1$ and $B_2$ given by
\beqa  B_1(f,g) &=& b^{\mu\nu}\partial_\mu f\partial_\nu g 
\cr & &\cr  B_2(f,g) &=& b_1^{\mu\nu}\partial_\mu f\partial_\nu g \cr &
&\cr &+ &
 \biggl(b^{\mu\nu\rho}\; +\; \frac 16(2 b^{\mu \sigma} + b^{\sigma\mu})
    \partial_\sigma b^{[\nu\rho]}
\; -\;\frac 12 b^{ \sigma\rho} \partial_\sigma b^{\mu\nu}\biggr) 
 \partial_{\mu,\nu} f\partial_\rho
    g \cr & &\cr&  +&  \biggl(b^{\mu\nu\rho}\; -\; \frac 16(
    b^{\mu \sigma} +2 b^{ \sigma\mu}) \partial_\sigma b^{[\nu\rho]} \; -\;\frac 12 b^{\rho \sigma} \partial_\sigma b^{\mu\nu}\biggr) 
 \partial_\rho f\partial_{\mu,\nu}
    g\cr & &\cr&  +&  b^{\mu\nu\rho\sigma}( \partial_{\mu ,\nu,\rho}f \partial_\sigma
    g +\partial_\sigma f \partial_{\mu,\nu,\rho}g)\;  +\;\frac
    {1}2( 3 b^{\mu\nu\rho\sigma}+ b^{\mu\rho} b^{\nu\sigma})
    \partial_{\mu,\nu} f\partial_{\rho,\sigma}
    g  \;,\cr & &
\label{sognrlstr}  \eeqa
where the tensors $
b^{\mu\nu...\rho}$ are expressed in terms of $\alpha^{\mu\nu}$ and  $
s^{\mu\nu...\rho}$ according to 
\beqa b^{\mu}&=& s^{\mu} \cr & &\cr b^{\mu\nu}&=&\frac i2 \alpha^{\mu\nu} + s^{\mu\nu} \cr & &\cr  b_1^{\mu\nu}&=&\frac i2
 T^{(s)}_1 \alpha^{\mu\nu}-s^\mu
s^\nu -\Bigl(\frac i2\alpha^{\mu\sigma} +
s^{\mu\sigma}\Bigr)\partial_\sigma s^\nu -\Bigl(\frac i2\alpha^{\sigma\nu} +
s^{\sigma\nu}\Bigr)\partial_\sigma s^\mu\cr & &\cr
 b^{\mu\nu\rho}&=&s^{\mu\nu\rho}-\frac 14 s^{(\mu\nu} s^{\rho)
  }\cr & &\cr b^{\mu\nu\rho\sigma}&=&s^{\mu\nu\rho\sigma}-\frac 1{48} s^{(\mu\nu} s^{\rho
  \sigma)}\;,
\label{ggtrns}\eeqa and  $b^{[\mu\nu]}=b^{\mu\nu}-b^{\nu\mu}$. We
introduced the vector  field $b^\mu$ for the sake of 
completeness.  Although
 it doesn't appear directly in the star product, it does appear in the
 measure. [See eq. (\ref{gnrlmsrscd}) below.]
 (\ref{sognrlstr}) reduces to (\ref{strknstvh})
when $ s=0$,
corresponding to the Kontsevich gauge.  
The star commutator is now
\beqa [f,g]_{\star_B} &=&\hbar \;(b^{[\mu\nu]}+\hbar b_1^{[\mu\nu]})\partial_\mu f\partial_\nu g \; +\;\frac
    {\hbar^2}2(  b^{\mu\rho} b^{\nu\sigma}- b^{\rho\mu} b^{\sigma\nu})
    \partial_{\mu,\nu} f\partial_{\rho,\sigma}g
\cr & &\cr&  +&
\; 
    \frac{\hbar^2}2 \biggl( b^{(\mu \sigma)}
    \partial_\sigma b^{[\nu\rho]}+  b^{[ \rho\sigma]}
 \partial_\sigma b^{\mu\nu}\biggr) (
 \partial_{\mu,\nu} f\partial_\rho
    g-  \partial_\rho
    f\partial_{\mu,\nu} g) \;  + \;{\cal O}(\hbar^3) \;.  \eeqa
As in (\ref{cmtvlmt}), the leading term  is  $ i\hbar \; \{ f,g \}$.

A subsequent gauge transformation can be performed  using $\Lambda$ in
(\ref{lmbdauxpns})   to map the star product $\star$ to $\star'$.  Now
write 
$\Lambda_k=T_k^{(\lambda)}$, with $T_k^{(\lambda)}$ given by
(\ref{tausubn}) and $\lambda $ denoting symmetric tensors
$\lambda=(\lambda^{\mu_1},\lambda^{\mu_1\mu_2},\lambda^{\mu_1\mu_2\mu_3},...)$.
  $\star'$ is again of the form  (\ref{gnrlfrmnqc}), with   $
b^{\mu\nu...\rho}$ in (\ref{sognrlstr}) replaced by   $
b^{'\mu\nu...\rho}$ defined by
\beqa
  b^{\mu}&\rightarrow & 
b'^{\mu}=b^{\mu}+
\lambda^{\mu}\cr & &
\cr b^{\mu\nu}&\rightarrow & 
b'^{\mu\nu}=b^{\mu\nu}+
\lambda^{\mu\nu}\cr & &
\cr   b_1^{\mu\nu}&\rightarrow & 
b_1^{'\mu\nu}=b_1^{\mu\nu}
\;+\; T^{(\lambda)}_1
 b^{\mu\nu}-\lambda^\mu
\lambda^\nu\; -\;b^{\mu\rho}\partial_\rho
\lambda^\nu-b^{\rho\nu}\partial_\rho \lambda^\mu-\lambda^{\rho(\mu}\partial_\rho
\lambda^{\nu )}\cr & &\cr b^{\mu\nu\rho}&\rightarrow & 
b'^{\mu\nu\rho}=b^{\mu\nu\rho}+
\lambda^{\mu\nu\rho}+\frac 14 \Bigl(
\lambda^{\sigma(\mu}\partial_\sigma b^{\nu\rho )}- \lambda^{(\mu\nu}\lambda^{\rho
  )}\Bigr)\cr & &\cr  b^{\mu\nu\rho\sigma}&\rightarrow & 
b'^{\mu\nu\rho\sigma}=b^{\mu\nu\rho\sigma}+
\lambda^{\mu\nu\rho\sigma}-\frac 1{48}
 \lambda^{(\mu\nu}\lambda^{\rho\sigma
  )}\;.
\label{sosubggtrns}\eeqa 
A gauge invariant antisymmetric tensor can be constructed from the
fields  $
b^{\mu\nu...\rho}$.  It is simply the zeroth order term  in an  $\hbar$ expansion expression for
$\alpha^{\mu\nu}$ re-expressed in terms of   $
b^{\mu\nu...\rho}$.  We remark that  the subset of inner automorphisms (\ref{inrot}) are
contained in the group generated by  (\ref{tausubn}).  For example, in that case
$T_1=2b^{\mu\nu} \lambda^{-1}_\star \partial_\mu\lambda \;\partial_\nu$.

\subsection{Measure}

The measure $d\mu_\star$  associated with the above star product
should satisfy   (\ref{trofmsrfrmk})  (for functions $f_0$ vanishing
sufficiently rapidly at infinity) and  reduce to 
$ d\mu_{0} =  d^{2d}x\;\Omega_0(x)$  in the
commutative limit. 
If  we expand $d\mu_{\star}$ in $\hbar$, \be d\mu_{\star}= d^{2d}x\;\Bigl(\Omega_0  \;+\; \hbar \Omega_1  \;+\;{\cal
  O}(\hbar^2)\;\Bigr)\;.\ee 
 $\Omega_0$ is gauge invariant.  We can substitute into
 (\ref{nvrsftrc}) to obtain the gauge transformations
 of the higher order corrections to the measure.
For example, if under  the action of $\Lambda$, $\Omega_1$ goes to
$\Omega'_1$ then
\be \int  d^{2d}x\;(\Omega'_1 -\Omega_1) f = - \int  d^{2d}x\;\Omega_0\;
T_1^{(\lambda)} f\;. \ee
Upon integrating by 
parts and assuming functions  $f$ vanish
sufficiently rapidly at infinity  one gets
\be  \Omega'_1 =\Omega_1  \;+\;
\partial_\mu (\Omega_0 \lambda^{\mu})
- \frac 1 2 \;
\partial_{\mu,\nu} (\Omega_0 \lambda^{\mu\nu}) \;.
\ee

\subsection{Connection and curvature}

For simplicity, here and in the following section, we work in the of equivalence class   $\{\star\}_\theta $ containing the Groenewold-Moyal star product.
Now  write the differential operator-valued potentials $A_{k,\mu}$ in
(\ref{almbdauxpns}) according to
 $A_{k,\mu}=T_k^{(a_\mu)}$  given in
(\ref{tausubn}), where here $a_\mu$ denote the tensors
$a_\mu=(a_\mu^{\mu_1},a_\mu^{\mu_1\mu_2},a_\mu^{\mu_1\mu_2\mu_3},...)$.
From (\ref{gtfdvptnls}), using (\ref{sdpespps}) and (\ref{invsdpespps}), we then deduce the following gauge
transformations for $a_\mu^{\rho\sigma...\eta}$:
\beqa a_\mu^{\rho}\;\;\;\rightarrow\;\;\;\; a_\mu^{'\rho}&=& (a_\mu-\partial_\mu\lambda)^\rho +
\hbar\; \Bigl(T_1^{(\lambda)} a_\mu^{\rho}-
T_1^{(a_\mu-\partial_\mu\lambda)}\lambda^\rho\Bigr) +  {\cal
  O}(\hbar^2)\cr & &\cr
a_\mu^{\rho\sigma}\;\;\rightarrow \;\;  a_\mu^{'\rho\sigma}&=& (a_\mu-\partial_\mu\lambda)^{\rho\sigma} +
\hbar\; \Bigl(T_1^{(\lambda)} a_\mu^{\rho\sigma}-
T_1^{(a_\mu-\partial_\mu\lambda)}\lambda^{\rho\sigma}\cr &
&\cr& &\qquad\qquad +\;\frac 12\partial_\mu(\lambda^{(\rho}\lambda^{\sigma
  )}) -
(a_\mu-\partial_\mu\lambda)^{\xi(\rho}\partial_\xi\lambda^{\sigma)}+
\lambda^{\xi(\rho}\partial_\xi a^{\sigma)}_\mu \Bigr) +  {\cal
  O}(\hbar^2)\cr & &\cr
a_\mu^{\rho\sigma\eta}\;\rightarrow \;  a_\mu^{'\rho\sigma\eta}&=& (a_\mu-\partial_\mu\lambda)^{\rho\sigma\eta} +
\frac 14\; \Bigl( \lambda^{\xi(\rho} \partial_\xi (a_\mu-\partial_\mu\lambda)^{\sigma\eta)}   -
(a_\mu-\partial_\mu\lambda)^{(\rho}\lambda^{\sigma\eta)}   \cr &
&\cr& &\qquad\qquad +\;\partial_\mu(\lambda^{\xi(\rho}\partial_\xi\lambda^{\sigma\eta  )})\;\Bigr) +  {\cal
  O}(\hbar)\cr & &\cr
a_\mu^{\rho\sigma\eta\xi}\rightarrow 
a_\mu^{'\rho\sigma\eta\xi}&=&
(a_\mu-\partial_\mu\lambda)^{\rho\sigma\eta\xi} +\frac 1{48} \partial_\mu(\lambda^{(\rho\sigma}\lambda^{\eta\xi  )}) +  {\cal
  O}(\hbar)\;. \eeqa
Using (\ref{fsntcs}) we can construct the field strengths
$F_{k,\mu\nu}=T_k^{(f_{\mu\nu})}$ where here $f_{\mu\nu}$ denotes the
tensors $f_{\mu\nu}^{\mu_1},f_{\mu\nu}^{\mu_1\mu_2},f_{\mu\nu}^{\mu_1\mu_2\mu_3},...$,
\beqa f_{\mu\nu}^{\rho}&=& \partial_\mu a^\rho_\nu +\hbar
T_1^{(a_\mu)}a_\nu^\rho - (\mu\rightleftharpoons \nu)  +  {\cal
  O}(\hbar^2)\cr & &\cr f_{\mu\nu}^{\rho\sigma}&=& \partial_\mu a^{\rho\sigma}_\nu +\hbar
\Bigl(T_1^{(a_\mu)}a_\nu^{\rho\sigma} + a^{\xi(\rho}_\mu\partial_\xi a^{\sigma)}_\nu\Bigr) - (\mu\rightleftharpoons \nu)  +  {\cal
  O}(\hbar^2)\cr & &\cr f_{\mu\nu}^{\rho\sigma\eta}&=& \partial_\mu
a^{\rho\sigma\eta}_\nu + \frac 14
a^{\xi(\rho}_\mu\partial_\xi a^{\sigma\eta)}_\nu - (\mu\rightleftharpoons \nu)  +  {\cal
  O}(\hbar)\cr & &\cr f_{\mu\nu}^{\rho\sigma\eta\xi}&=& \partial_\mu
a^{\rho\sigma\eta\xi}_\nu  - (\mu\rightleftharpoons \nu)  +  {\cal
  O}(\hbar)\;. \eeqa
They gauge transform according to 
\beqa f_{\mu\nu}^{\rho}\;\;\;\rightarrow\;\;\;\; f_{\mu\nu}^{'\rho}&=&
f_{\mu\nu}^{\rho}  +
\hbar\; \Bigl(T_1^{(\lambda)}  f_{\mu\nu}^{\rho}-
T_1^{(f_{\mu\nu})}\lambda^\rho\Bigr) +  {\cal
  O}(\hbar^2)\cr & &\cr
f_{\mu\nu}^{\rho\sigma}\;\;\rightarrow \;\;\;  f_{\mu\nu}^{'\rho\sigma}&=& f_{\mu\nu}^{\rho\sigma} +
\hbar\; \Bigl(T_1^{(\lambda)}f_{\mu\nu}^{\rho\sigma}-
T_1^{(f_{\mu\nu})}\lambda^{\rho\sigma} +\lambda^{\xi(\rho}\partial_\xi
f_{\mu\nu}^{\sigma)} - f_{\mu\nu}^{\xi(\rho}\partial_\xi\lambda^{\sigma)} \Bigr) +  {\cal
  O}(\hbar^2)\cr & &\cr
f_{\mu\nu}^{\rho\sigma\eta}\;\rightarrow \;  f_{\mu\nu}^{'\rho\sigma\eta}&=& f_{\mu\nu}^{\rho\sigma\eta} +
\frac 14\; \Bigl( \lambda^{\xi(\rho} \partial_\xi
f_{\mu\nu}^{\sigma\eta)}
-f_{\mu\nu}^{\xi(\rho}\partial_\xi\lambda^{\sigma\eta)}  \;\Bigr) +  {\cal
  O}(\hbar)\cr &
&\cr f_{\mu\nu}^{\rho\sigma\eta\xi}\rightarrow f_{\mu\nu}^{'\rho\sigma\eta\xi}&=&
f_{\mu\nu}^{\rho\sigma\eta\xi} +  {\cal
  O}(\hbar)\;. \eeqa

\subsection{Field equations}

We can now substitute the above expansion for the field strength
tensor into the sourceless Yang-Mills type equation  (\ref{ymteq})
\beqa && \partial^\mu f^\rho_{\mu\nu} +\hbar
T_1^{(a^\mu)} f^\rho_{\mu\nu} +  {\cal
  O}(\hbar^2)=0\cr & &\cr && \partial^\mu f_{\mu\nu}^{\rho\sigma}  +\hbar
\Bigl(T_1^{(a^\mu)} f_{\mu\nu}^{\rho\sigma} +
[a^\mu]^{\xi(\rho}\partial_\xi  f^{\sigma)}_{\mu\nu}\Bigr)  +  {\cal
  O}(\hbar^2)=0\cr & &\cr & & \partial^\mu  f_{\mu\nu}^{\rho\sigma\eta}
  + \frac 14
[a^\mu]^{\xi(\rho}\partial_\xi  f_{\mu\nu}^{\sigma\eta)} +  {\cal
  O}(\hbar)= 0\cr & &\cr & & \partial^\mu  f_{\mu\nu}^{\rho\sigma\eta\xi}
  +  {\cal
  O}(\hbar)=0\;,\label{slymfeq} \eeqa where $a^\mu$ is obtained from $a_\mu$ assuming a
flat metric.

From the action (\ref{gnvactnfrmtr}) we can also easily obtain the field
equation for the scalar field $\phi$ in a background $A_\mu$.  For
simplicity, choose $\star$ to be the Groenewold-Moyal star.  Variations
of $\phi$ lead to an exactly conserved current
\be \partial_\mu k^\mu =0\;,\ee
where
\be  k^\mu = D[A]^\mu\phi + \hbar\; a^\mu_\rho  D[A]^\rho\phi -\frac \hbar 2
\partial_\rho (a^{\mu\rho}_\sigma   D[A]^\sigma\phi ) +  {\cal
  O}(\hbar^2)\;.\ee
One recovers the commutative result $\partial_\mu\partial^\mu\phi=0$
 at zeroth order in $\hbar$.
On the other hand, in order to couple  $\phi$ to the above gauge
theory   one should search currents $J^\mu$ which are
covariantly conserved, i.e.  (\ref{nftcnsrvlws}),  and gauge transform
as (\ref{gtrnofJ}).  Let us assume  they exists and can be expanded as in 
(\ref{drvldcrnt}), with $J_{k,\mu}$  given by $J_{k,\mu}=T_k^{(j_\mu)}$. $j_\mu$ denote the tensors
$j_\mu=(j_\mu^{\mu_1},j_\mu^{\mu_1\mu_2},j_\mu^{\mu_1\mu_2\mu_3},...)$,
which now enter on the right hand sides of (\ref{slymfeq}).
From (\ref{nftcnsrvlws}),
$T_1^{(j_\mu)}$ is exactly conserved at zeroth order in $\hbar$.
Candidates for first two
currents (up to an overall
factor) are
\beqa j^\rho_\mu&=& \partial_\mu\phi \partial^\rho\phi - \phi
\partial_\mu \partial^\rho\phi +  {\cal
  O}(\hbar)\cr & &\cr  j^{\rho\sigma}_\mu&=& \partial_\mu\phi \partial^{\rho,\sigma}\phi - \phi
\partial_\mu \partial^{\rho,\sigma}\phi +  {\cal
  O}(\hbar)\;.
\eeqa   The next order  expressions for these currents will contain
the potentials $a^\mu$ and they are expected to be nonlocal.

\section{Conclusion} 
\setcounter{equation}{0}

In the previous sections we developed tools for writing gauge theories on the space
$\{\star\}_\alpha$ of equivalent star products associated with any
given Poisson bi-vector $\alpha$.    The gauge theories can be regarded
as an extension of noncommutative $U(1)$ gauge theory.  Since general
gauge transformations induce ${\cal O}(\hbar)$ corrections in the
integration measure, they cannot be regarded as purely internal
transformations.

Although it is not difficult to write down matter field actions, as
for example in (\ref{gnvactnfrmtr}), a final ingredient is needed  in
order to introduce kinetic terms for the
infinitely many gauge fields in $F_{\mu\nu}$, namely the trace Tr over differential operators.  In additional to satisfying the usual
trace property, Tr$\;F_{\mu\nu}F^{\mu\nu}$ should reduce to the usual
action for  noncommutative $U(1)$ gauge fields upon restricting $F_{\mu\nu}$ to (\ref{ncuonef}).  Other familiar
noncommutative field theories may be contained in the full action.
In order to make contact
with physical theories,  mechanisms, such as the Higgs mechanism,
should  be applied to give (large)
masses to all but a finite number of the gauge fields.  This may then
involve  introducing additional derivative-valued fields. 

Finally, a more ambitious project would be to write down field theories
on the space of {\it all} equivalence classes $\{\star\}_\alpha$ of
star products.  This means  making the Poisson bi-vector $\alpha$  dynamical,
and then as a result all of the  bi-differential operators
$B_n$ in the star product (\ref{gnrlfrmnqc}) dynamical.
Variations of these operators must then include diffeomophisms on the
underlying manifold, at all orders in $\hbar$, setting the possible framework for a quasi-classical
approximation to quantum gravity.\cite{Doplicher:1994tu}

\bigskip

\noindent
{\bf Acknowledgment}

\noindent
The work of A.P. has been supported by FAPESP grant number 06/56056-0.

\end{document}